\documentclass[aps,epsf,preprint]{revtex4}
\usepackage{graphicx}% Include figure files
\usepackage{bm}% bold math
\begin{document}
\newcommand{\Sv}{{\bm S}}
\newcommand{\eF}{{\epsilon_F}}
\newcommand{\kF}{{k_F}}
\newcommand{\sigmav}{{\bm \sigma}}
\newcommand{\tauphi}{{\tau_\varphi}}%
\newcommand{\ellphi}{{\ell_\varphi}}%
\newcommand{\ellphip}{{\ell_\varphi}'}%
\newcommand{\phitil}{\frac{\phi}{\phi_0}}%
\newcommand{\xv}{{\bf x}}
\newcommand{\rv}{{\bf r}}
\newcommand{\kv}{{\bf k}}
\newcommand{\qv}{{\bf q}}
\newcommand{\Vv}{{\bf V}}
\newcommand{\Mv}{{\bf M}}
\newcommand{\Av}{{\bf A}}
\newcommand{\Bv}{{\bf B}}
\newcommand{\av}{{\bf a}}
\newcommand{\ra}{{\rightarrow}}
\newcommand{\DOS}{{\nu}}
\newcommand{\kB}{{k_B}}
\newcommand{\Area}{A_{\rm r}}
\newcommand{\DOStD}{\DOS_{2D}}
\newcommand{\Ha}{H_{a}}
\newcommand{\zeeman}{h}
\newcommand{\gap}{\Delta_g}

\title{
On Aharonov-Bohm oscillation in a ferromagnetic ring
} 
\author{Gen Tatara}
\affiliation{
Graduate School of Science, Osaka University, Toyonaka Osaka 560-0043, 
Japan}
\author{Hiroshi Kohno} 
\affiliation{
Graduate School of Engineering Science, Osaka University, 
Toyonaka Osaka 560-8531, Japan}
\author{Edgar Bonet}
\author{Bernard Barbara}
\affiliation{
CNRS - Laboratoire de Magn\'etisme Louis N\'eel,
25 Ave. des Martyrs, BP 166, 
38042, Grenoble Cedex 09, France
}

\date{\today}
\begin{abstract}
 Aharonov-Bohm effect in a ferromagnetic thin ring in diffusive regime is theoretically studied by calculating the Cooperon and Diffuson.
In addition to the spin-orbit interaction, we include the spin-wave excitation and the spin splitting, which are expected to be dominant sources of dephasing in ferromagnets at low temperatures. The spin splitting turns out to kill the spin-flip channel of Cooperon but leaves the spin-conserving channel untouched. 
For the experimental confirmation of interference effect (described by Cooperons) such as weak localization and Aharonov-Bohm oscillation with period $h/2e$, we need to suppress the dominant dephasing by orbital motion.
To do this we propose experiments on a thin film or thin ring with magnetization and external field perpendicular to the film, in which case the effective field inside the sample is equal to the external field (magnetization does not add up).
The field is first applied strong enough to saturate the magnetization and then carrying out the measurement down to zero field keeping the magnetization nearly saturated, in order to avoid domain formations (negative fields may also be investigated if the coercive field is large enough).  
\end{abstract}
\maketitle
%%%%%%%%%%%%%%%%%%%%%%%%%%%%%%%%%%%%%%%%%%%%%%
Quantum electron transport in mesoscopic systems has been intensively studied for this couple of decades.
Typical phenomena are weak localization\cite{Bergmann84,Lee85} and Aharonov-Bohm oscillation\cite{Aronov87}, both arising from the interference of electron wave function.
Many experiments have been carried out on various non-magnetic metals and semiconductors.
Although the effect of magnetic impurities in non-magnetic hosts has been studied in detail\cite{Bergmann84}, ferromagnetic metals themselves have not been explored in the context of quantum transport until very recently.
One of the reasons may be that the dephasing mechanism in ferromagnetic metals was believed to be much more efficient and complex than non-magnetic cases with, as a result, complete destruction of interferences.
However, these apparent disadvantages are not always crucial, as discussed briefly in Ref. \cite{TB01}, where it was indicated that Aharonov-Bohm effect should be observed in ferromagnets. 
The first complexity is the existence of the internal field (related to the magnetization), $M$, 
which generally can lead to dephasing even in the absence of the applied magnetic field.
The field depends much on sample shape. Here, we consider an ultra-thin ring with sufficiently high perpendicular anisotropy so that the magnetization is perpendicular to the ring.
In this case, the total field in the sample is simply equal to the external field, $B_0=\mu_0 H$. In fact, the effective field inside the sample is $B=B_0+M+B'$, where 
$B'\equiv \mu_0 H_D$ denotes the field produced by the surface magnetic charge. 
The total field $\Bv$ satisfies the Maxwell equation, $\nabla\cdot \Bv=0$, and thus its component perpendicular to the plane is continuous across the surface of the ring; 
i.e., $B=B_0$, or $B'=-M$. 
Hence the effect of $M$ on the orbital motion can be neglected.
Second, the spin splitting due to effective exchange interaction with the local spin, which arises from the $s$-$d$ mixing, needs to be taken account of as a source of dephasing. 
The splitting turns out to suppress spin flip channel of Cooperon and Diffuson but spin-conserving channels survive. Here Cooperon is a particle-particle propagator, which represents the interference effect, and Diffuson is the particle-hole propagator representing the diffusive motion\cite{Altshuler85}.
Third, ferromagnets generally contain domains and thus it is not always
easy to identify magnetic structure. 
In addition, domain walls may also cause dephasing\cite{TF97}.
Dephasing due to domain structures can be easily avoided by applying a magnetic field larger than the saturation field, $H_s$.
In nanostructures with strong perpendicular anisotropies, $H_s \sim H_c$, the coercive field.
Hysteretic behavior of ferromagnets with large perpendicular anisotropies should allow to reduce the external field without affecting the magnetic state, down to zero and even to negative values $(H>-H_c)$. 
From this point of view, hard magnets with very sharp (square-like) hysteresis loops would be suitable to study electronic coherence in ferromagnets.

The first experimental study of the Aharonov-Bohm effect in ferromagnetic metals was 
carried out quite recently\cite{Kasai02}, where the Aharonov-Bohm oscillation was observed on a permalloy ring in the presence of an applied field of $\gtrsim 3 $T. 
The Fourier transform of the conductance exhibits a peak corresponding to $h/e$ oscillation. This oscillation period seems to be due to the interference of a single electron propagator\cite{Aronov87}.
The oscillation period of $h/(2e)$ (called Altshuler-Aronov-Spivak oscillation\cite{Altshuler81}) was not seen. 
We believe this could be because of external field $\sim 3$T, which kills the Cooperon (see below) (In non-magnetic case, similar vanishing of $h/2e$ oscillation by applied field was observed\cite{Umbach86}).
Although the dephasing length was estimated there to be $\sim 5000\AA$\cite{Kasai02}, this appears to be too long considering the effect of the external field of $\sim3$T, which makes the length shorter than $L_B\equiv \sqrt{12}(\hbar/eaB_0)\sim 200\AA$ (see eq. (\ref{ellphipdef})) for a ring of width $a=400\AA$. Such long dephasing length of $5000\AA$ appears also to be inconsistent with observation of no $h/2e$ oscillation. 
In the case of Ni wires, the resistivity measured down to 20mK was shown to be explained by the enhanced electron-electron interaction due to the diffusive motion\cite{Ono01}, which indicates that Diffuson (particle-hole) channel exists, but no clear sign of Cooperon channel was observed.
Thus, at present, although experiments suggest the existence of Diffuson in ferromagnets, there is no indication of the electron coherence represented by Cooperons.

The aim of this paper is to give a thorough description of Aharonov-Bohm effect  in a ferromagnetic thin ring in diffusive regime, by calculating the Cooperon and Diffuson.
In addition to the spin-orbit interaction, we include the spin-wave excitation and the spin splitting, which are expected to be dominant sources of dephasing in ferromagnets at low temperatures. 
Spin-flip scattering by single localized spins (similar to those of magnetic impurities in non-magnetic metals) must be suppressed by the strong exchange interaction in the ferromagnet. 
Instead of that, spin-wave excitation with long wavelength would be important.
In fact, spin waves turn out to result in strong dephasing effect if gapless.
In reality, this effect should be suppressed by perpendicular magnetic anisotropy 
and/or 
by the application of an external field $B_0$ (as well as finite system size).
In zero field, the spin-wave gap depends on crystal-field symmetry and parameters. In a first approximation we will simply assume that it is given by 
%$\gap = M_s(\mu_0\Ha + B_0)$,  
%where $\Ha$ is the anisotropy field and $M_s$  the spontaneous magnetization. 
$\gap = \hbar\gamma(\Ha + B_0/\mu_0)$,  
where $\Ha$ is the anisotropy field and $\gamma$ is the gyromagnetic ratio. 
Dephasing due to spin wave 
can be neglected if $\kB T \ll \gap$, and therefore the effect could be tuned from sample to sample by changing the anisotropy energy, or on a given sample by changing the applied field. 
This will allow to study dephasing effects by spin waves, if the $h/(2e)$ oscillations (Altshuler-Aronov-Spivak oscillations) were to be observed in a ferromagnet. We believe that this will be the case because, contrary to previous experiments\cite{Kasai02,Ono01}, we suggest low-field experiments in ultra-thin ferromagnets with perpendicular anisotropy. In this case, as we argued above, the field coherence length $L_B$ should not be affected by the magnetization $M_s$ of the ferromagnet, and therefore should be as large as in non-magnetic material. 
In low applied field $B_0$,   $L_B \propto 1/B_0$  should  be long enough so that $h/2e$ oscillation (Cooperon) would be observed.

We consider a ring with mean radius $R$ and width $a$ 
(i.e., the outer and inner radii are $R+\frac{a}{2}$ and
$R-\frac{a}{2}\equiv R_0$, respectively), and thickness $b$.
The width and thickness are assumed to be smaller than the coherence
length of the electron, $\ell_\varphi$, and $R$.
We consider the equation of motion of the Cooperon in this ring, but first without spin-orbit interaction, spin waves and spin splitting\cite{Aronov87}
\begin{equation}
\left[ D(-i\nabla-2e\Av)^2+\frac{1}{\tauphi}\right]C_0(\rv-\rv')=\delta^3(\rv-\rv').
\label{eqofmo}
\end{equation}
Here $D=(k_F/m)^2(\tau/3)$ is the diffusion constant ($\tau$ is the elastic lifetime)
and $\tauphi$ ($\sqrt{D\tauphi}\equiv \ell_\varphi$) is the dephasing time
due to inelastic scattering from non-magnetic sources, e.g.,
electron-electron interaction and phonons.
The applied magnetic field $\Bv_0$ is in $z$-direction,i.e.,
 perpendicular to the ring.
We consider the case where the magnetization, $\Mv$, is also perpendicular to the ring and constant inside the ring.
It is important to note here that in this configuration with a thin ring, the total field, $B\equiv B_0+M+B'$ is identical outside and inside the ring, since $\nabla\cdot\Bv=0$ requires the continuity of $B$ (in other words,
the field $B'$ due to surface magnetic charges cancels the effect of $M$). 
Thus $B=B_0$ and the vector potential is identical to that in the non-magnetic case;
$\Av=\frac{B_{0}}{2}(-y,x,0)$.
The equation is thus rewritten as
\begin{equation}
\left[ \partial_r^2+\frac{1}{r}\partial_r +\frac{1}{r^2} \partial_\theta^2
-\frac{2ieB_{0}}{\hbar} \partial_\theta -\left(\frac{eB_{0}}{\hbar}\right)^2 r^2
-\ellphi^{-2}\right]C_0(\rv-\rv')=-\frac{1}{D}\delta^3(\rv-\rv')
\label{eq2}
\end{equation}
where $\tan\theta\equiv y/x$.
The boundary condition in $r$-direction is given by 
$\frac{\partial C_0}{\partial r}|_{r=R\pm\frac{a}{2}} =0$ (open boundary).
Since we assume $a\ll\ellphi$, only uniform mode contributes in $r$-direction.
The equation for this uniform mode is obtained by integrating 
eq. (\ref{eq2}) over $r$ from $R-\frac{a}{2}$ to $R+\frac{a}{2}$, for instance, 
$r^2 \rightarrow  \langle r^2 \rangle\equiv \frac{1}{a}
\int_{R-\frac{a}{2}}^{R+\frac{a}{2}}dr r^2= R^2 +\frac{a^2}{12}$.
The equation thus reduces to
\begin{equation}
\left[ \partial_u^2
-4i\frac{\phi}{R\phi_0}\partial_u -4\left(\frac{\phi}{R\phi_0}\right)^2
-{\ellphip}^{-2}\right]C_0(u-u')=-\frac{1}{D ab}\delta(u-u')
\label{eq3}
\end{equation}
where $u\equiv R\theta$, $\phi\equiv \pi R^2B_0$ 
($\phitil= eB_{0}R^2/(2\hbar)$, and 
$\phi_0\equiv h/e$ is the flux quantum).
The dephasing length with the effect of the orbital motion caused by $B_0$ is thus given by the same expression without $M$ as in non-magnetic case\cite{Aronov87}
\begin{equation}
\ellphip^{-2}\equiv \ellphi^{-2}+\frac{1}{12}\left(\frac{eaB_0}{\hbar}\right)^2
\label{ellphipdef}.
\end{equation}
%[\frac{1}{12}(B_0+M)^2+\frac{1}{2}M(B_0+M)\frac{R-\frac{a}{2}}{R+\frac{a}{2}}
%+\frac{1}{4}M^2(\frac{R-\frac{a}{2}}{R+\frac{a}{2}})^2$].
The on-site amplitude of the Cooperon (without spin-flip scattering), 
$C_0(0)$, is thus obtained as
\begin{equation}
C_0(0)=\frac{1}{\pi abL D} \sum_{\ell=-\infty}^\infty
\frac{1}{ \frac{1}{R^2} \left({\ell}-2\phitil\right)^2
+\ellphip^{-2}},
\end{equation}
where $\ell$ runs over integers.

We now include the spin-orbit coupling, scattering by spin waves, and spin splitting;
\begin{equation}
H'\equiv 
\sum_{\kv,\kv'}i\lambda_{so}(\kv\times\kv')\cdot 
c^\dagger_{\kv'} \sigmav c_\kv
%+\sum_{\kv,\kv'}v_s \Sv_{\rm imp}\cdot c^\dagger_{\kv'} \sigmav c_\kv
+\sqrt{2S}J \sum_{\qv,\kv}\sum_{\pm} a_\qv^{\pm} 
c^\dagger_{\kv+\qv} \sigma_{\pm} c_\kv
+gM\sum_\kv c^\dagger_\kv\sigma_z c_\kv.
\label{hp}
\end{equation}
The scattering by spin waves is represented by the second term, where
$a^+\equiv a^\dagger$ and $a^{-}\equiv a$ are spin-wave operators.
Assuming low temperature, the spin-wave interaction is included only at the 
linear order.
The last term represents the spin splitting proportional to the magnetization.
The spin flip processes by spin waves result in new channels in Cooperon\cite{Hikami80}
The spin splitting results in a dephasing in the total $S^z=0$ channel\cite{Maekawa81}.
The full Cooperon with $H'$ included is obtained 
as\cite{Hikami80,Aronov87,Maekawa81,Fukuyama85}
\begin{eqnarray}
C(0)\! &=& \!
\frac{1}{abL D} \sum_{\ell=-\infty}^\infty
\left[ 
\frac{1}{\frac{1}{R^2} \left({\ell}-2\phitil\right)^2 \!
+L_1^{-2}}
+\frac{ {1/2}(L_2^{-2}-L_3^{-2}) } 
{ \left[ \frac{1}{R^2} \left({\ell}-2\phitil\right)^2 \!
  +L_2^{-2} \right]  \!\!\!
  \left[ {\frac{1}{R^2} \left({\ell}-2\phitil\right)^2 \!
  +L_3^{-2}} \right] \!
  +4 L_M^{-4}   }
\right] ,
\nonumber\\
&&\label{C2}
\end{eqnarray}
where 
\begin{eqnarray}
L_1^{-2} &\equiv& \ellphi^{-2}+\frac{1}{12}\left(\frac{eaB_0}{\hbar}\right)^2
+\frac{1}{D}\left( 
\frac{1}{\tau_{so}^z}+\frac{1}{\tau_{so}^x}+\frac{1}{\tau_{sw}^x}
\right),
\nonumber\\
L_2^{-2} &\equiv& \ellphi^{-2}+\frac{1}{12}\left(\frac{eaB_0}{\hbar}\right)^2
+\frac{1}{D}
%\left( \frac{2}{\tau_{sw}^z}+
\frac{4}{\tau_{sw}^x}
%\right)
,
\nonumber\\
L_3^{-2} &\equiv& \ellphi^{-2}+\frac{1}{12}\left(\frac{eaB_0}{\hbar}\right)^2
+\frac{1}{D}
%\frac{2}{\tau_{sw}^z}+
\frac{4}{\tau_{so}^x},
\label{Lsdef}
\end{eqnarray}
and $L_M^{-2}\equiv gM/D$.
Here 
$1/\tau_{so}^\mu \equiv 2\pi \DOS \lambda_{so} ^2
\langle (\kv'\times\kv)_\mu^2 \rangle $, where $\mu=x,y,z$,
bracket denotes the average over configuration, $\DOS$ is the density of states,
and 
$1/\tau_{sw}^\mu$ 
is the spin-flip rate due to spin wave.
We have
assumed that  $1/\tau_{so}^x=1/\tau_{so}^y$.
There is no $z$-component in spin-wave scattering at the present lowest-order calculation (see eq. (\ref{hp})).
Spin-flip 
%%%component, 
contribution, 
$1/\tau_{sw}^x(=1/\tau_{sw}^y)$, is obtained as 
\begin{equation}
\frac{1}{\tau_{\rm sw}^{x}} =2\pi SJ^2 \sum_{\pm} 
\sum_\qv\frac{1}{\sinh \beta\omega_q}\delta(\epsilon_{\kv+\qv}\pm\omega_q),
\label{swtaudef}
\end{equation}
where $\beta=1/(k_B T)$ and $\omega_q\equiv \gap+Aq^2$ is the spin-wave energy, $\gap$ and $A$ being the spin-wave gap and stiffness, respectively.
The gap is written in terms of the magnetic anisotropy energy, $\Ha$, and the external field, $B_0$, as $\gap=\hbar\gamma(\Ha+B_0/\mu_0)$. The spin-wave stiffness is roughly given as
$A\simeq J/k_F^2$, where $J$ is the exchange coupling between the localized spins. 
The integration over $\qv$ is treated as two-dimensional 
(this is allowed if $\kB T \ll J/(k_F b)^2$).
The integration is carried out as
\begin{equation}
\frac{1}{\tau_{\rm sw}^x} =\frac{ SJ^2}{\pi}m \Area \sum_{\pm} 
\int_{q_{2\pm}}^{q_{1\pm}} \frac{qdq}{\sinh\beta\omega_q} 
\frac{1}{\sqrt{(q_{1\pm}^2-q^2)(q^2-q_{2\pm}^2)}},
\end{equation}
where $q_{1\pm}\equiv k_F+\sqrt{k_F^2\pm2m\gap}$, 
$q_{2\pm}\equiv k_F-\sqrt{k_F^2\pm2m\gap}$ and $\Area$ is the area of the ring.
(We approximated the projection of three-dimensional Fermi wavelength onto 
the plane by  $k_F$.)
Since $J\simeq Ak_F^2\gg \gap$, the integral is dominated by the contribution 
from the region close to the lower limit. We thus obtain
\begin{equation}
\frac{1}{\tau_{\rm sw}^x}\simeq 4\pi S \frac{\DOStD J^2}{\sinh\beta\gap},
\label{swresult}
\end{equation}
where $\DOStD\equiv \DOS/(k_F b)$ is the two-dimensional density of states.
The effect of spin wave is thus different in two cases;
$\hbar\gamma \Ha \gg \kB T$ and $\hbar\gamma \Ha \ll \kB T$.
In the first case of strong anisotropy, the spin-wave excitation is
negligible if $\kB T \leq \hbar\gamma\Ha$,  irrespective of the external field.
In the second case with small anisotropy, the dephasing
by spin wave is controlled by the external field;
%%%
it is suppressed if $\kB T\lesssim  \hbar\gamma B_0/\mu_0$.
Hence in this case, the oscillation would be visible only at high field region and would vanish at small field $\hbar\gamma B_0/\mu_0 \lesssim \kB T$.
For the observation of Aharonov-Bohm oscillation, large anisotropy energy is of course favorable.

We note that spin waves can be extremely dangerous if gapless.
In fact,  in the gapless case, long-range ($q\sim0$) contribution in 
eq. (\ref{swtaudef}) is given by 
$\frac{1}{\tau_{\rm sw}^{x}} \propto  \sum_\qv\frac{1}{q^3}$, which diverges 
($\propto L$ in the two-dimensional case).
This is compared to the phonon case, where the contribution is finite due to the linear energy dispersion and an extra factor proportional to $q$ from coupling constants\cite{Abrikosov63}.
Divergence of $\frac{1}{\tau_{\rm sw}^{x}}$ indicates that the linear approximation breaks down, and more sophisticated calculation including higher-order contributions is needed
to treat the gapless case correctly.
Here we will not go further in this direction, since in reality there is generally a gap. 

%%%%%%%%%%%%%%%%%%%%%%%%%%%%%%%%%%%%%%%%%%%%%%

The Cooperon is directly related to the quantum correction to the conductivity as
$\Delta \sigma=-\frac{2}{\pi}e^2 D C(0)$\cite{Aronov87}.
Noting $\ell$ in eq. (\ref{C2}) runs over all integers, it is seen that the conductance $G\equiv ab\sigma$ oscillates as a function of magnetic flux $\phitil$ with the period of 
%%%$\phitil =1/2$, 
$\Delta\phi /\phi_0 =1/2$, 
%%i.e., 
or 
$\Delta B_0=\phi_0/(2\pi R^2)$.
Note that the magnetization $M$ does not affect this oscillation period.
The oscillation appears also in the conductance fluctuation as a function of magnetic field (or correlation function between different magnetic fields). The fluctuation is given as\cite{Aronov87}
\begin{equation}
\langle \Delta G(B_0) \Delta G(B_0+b_0)\rangle
\simeq \frac{48e^4}{\pi^3}\frac{Dab}{R} ( C'(0)+D'(0) ),
\end{equation}
where
$C'(0)$ is the Cooperon connecting electrons with different field ($B_0$ and $B_0+b_0$), which is defined by eq. (\ref{C2}), but with 
$\phi={\pi R^2}(B_{0}+\frac{b_0}{2} )$
and $L_i$'s defined by eq. (\ref{Lsdef}) with $B_0$ replaced by $B_0+b_0/2$.
Diffuson contribution, $D'(0)$, is similarly  
given by the right hand side of eq. (\ref{C2}), but with $\phi={\pi R^2} b_0$
and $L_i$'s defined by eq. (\ref{Lsdef}) with $B_0$ replaced by $b_0/2$.
(The Diffuson is a particle-hole propagator which carries zero electric charge, and so $D'(0)$ is not affected by the field $B_0$ coupled to the center-of-mass motion.)

Let us look into each dephasing mechanism in eq. (\ref{Lsdef}). 
The non-magnetic part, $\ellphi$, would be identical as in the non-magnetic systems, i.e., the contribution is mainly from the electron-electron interaction at low temperatures\cite{Altshuler85,Bergmann84} (say, $T\lesssim1$K).
Experimentally $\ellphi$ is estimated to be $1\sim 2\mu$m in Al and 
Ag\cite{Chandra85,Umbach86}, which is long enough for submicron rings.
%Note that electron-electron interactions should even be smaller in a ferromagnet 
%where conduction electron spins are partially polarized (Pauli exclusion principle).
The spin-orbit interaction may be different in magnetic case, but is estimated in Ag as 
$\ell_{so}=1/\sqrt{D\tau_{so}}\sim 0.47\mu$m\cite{Umbach86}.
Now turn to dephasing of magnetic origin.
The dephasing length due to orbital effect, $L_B\equiv \sqrt{12}(\hbar/eaB_0)$ (the last term of eq. (\ref{ellphipdef})) can be short for a strong field;
For $a=400\AA$ and $B_0=1$T, $L_B=570\AA$.
But  $L_B$ can be easily controlled to be long enough by choosing $B_0$ to be small.
In non-magnetic case, clear oscillation pattern is observed for $B_0\lesssim 0.02$T\cite{Umbach86}.
In ferromagnetic case, such small-field experiment must be done after saturating the magnetization by a strong field (in order to avoid domain formation).
The dephasing length due to spin splitting is given by  
$L_M=\kF^{-1}\sqrt{\frac{2}{3} \kF \ell\frac{\epsilon_F}{\Delta}}$.
This can be short; in dirty case of $\kF \ell\simeq 10\sim100$, even if the splitting of $s$ electron is $\frac{1}{100}$ times smaller than that of $d$-electron ($\Delta_d/\epsilon_F\sim O(0.1)$ and so $\Delta/\epsilon_F\sim 0.001$), we have 
$L_M\simeq \kF^{-1}\times(100\sim300)$.
So the splitting is one of the dominant sources of dephasing in ferromagnetic systems\cite{TF97}, and the spin flip channel (the last term in eq. (\ref{C2})) can be neglected.
Thus only the Cooperon in the vanishing total spin ($S^z=0$) channel survives, which is calculated as 
(after the summation over $\ell$)
\begin{equation}
C(0)\simeq
\frac{1}{2ab D} 
L_1 \frac{\sinh\frac{L}{L_1}}
{ \cosh\frac{L}{L_1}-\cos(4\pi\phitil) }.
\label{cresult}
\end{equation}
It is important to note here that due to $L_M$, the anti-localization effect, which is expected when the spin-orbit interaction is strong in non-magnetic metals, does not appear in ferromagnets. (Anti-localization arises from spin-flip processes given by the last term in eq. (\ref{C2}).)
The Cooperon which appears in the fluctuation, $C'(0)$, is given by\cite{Aronov87}
\begin{equation}
C'(0)\simeq
\frac{1}{2ab D} 
L_1' \frac{\sinh\frac{L}{L_1'}}
{ \cosh\frac{L}{L_1'}-\cos(4\pi\frac{\phi+\Delta\phi/2}{\phi_0}) },
\label{cpresult}
\end{equation}
where $\Delta\phi\equiv b_0\pi R^2$ is the flux due to the field difference, and
\begin{equation}
{L_1'}^{-2} \equiv \ellphi^{-2}+\frac{1}{12}\left(\frac{ea}{\hbar}\right)^2 (B_0+b_0/2)^2
+\frac{1}{D}\left( 
\frac{1}{\tau_{so}^z}+\frac{1}{\tau_{so}^x}+\frac{1}{\tau_{sw}^x}
\right).
\end{equation}
Similarly, Diffuson is obtained as
\begin{equation}
D'(0)\simeq
\frac{1}{2ab D} 
L_1'' \frac{\sinh\frac{L}{L_1''}}
{ \cosh\frac{L}{L_1''}-\cos(4\pi\frac{\Delta\phi/2}{\phi_0}) },
\label{dpresult}
\end{equation}
where
\begin{equation}
{L_1''}^{-2} \equiv \ellphi^{-2}+\frac{1}{12}\left(\frac{eab_0/2}{\hbar}\right)^2
+\frac{1}{D}\left( 
\frac{1}{\tau_{so}^z}+\frac{1}{\tau_{so}^x}+\frac{1}{\tau_{sw}^x}
\right).
\end{equation}

Now we assume that $\ellphi$ and $\ell_{so}$ are longer than the sample length, $L$.
We neglect the spin-wave scattering, assuming low temperatures; $\kB T\lesssim \gap$.
For a ring of $L=1.5\mu$m and $a=400\AA$, magnetic field $B_0$ as large as  $0.038$T 
%%%
(at which $L_B \sim L$) kills the Cooperons $C(0)$ and $C'(0)$. 
%%%because of small $L_B$.
Hence no Aharonov-Bohm oscillation appears in the conductance itself for $B_0\gtrsim 0.038$T. Only the fluctuation of conductance shows oscillation due to $D'(0)$.
In contrast, at very small field (after saturating the magnetization), all of $C(0), C'(0)$ and $D'(0)$ survives, and oscillation will be seen in both $G$ and 
$\langle \Delta G(B_0) \Delta G(B_0+b_0)\rangle$.

%%<
In ref. \cite{Kasai02}, the dephasing length was estimated to be $\sim 5000\AA$ under the field of $3\sim4 $T. This is too long if we consider it as dephasing length of Cooperon ($L_1$), since the orbital effect at $B_0=3$T would be strong enough to kill the coherence at the length of $L_B\sim 200\AA$.
In our opinion, the length above corresponds to the length scale of Diffuson, $L_1''$, which is not affected by $B_0$, and further studies seem to be needed to confirm the coherence represented by Cooperon.
%%>

In conclusion, we have shown that interference effects in conductance ($h/2e$ oscillation or the AAS effect) should be observed in ferromagnetic films or rings, provided the magnetization is perpendicular to the surface and the sample is thin enough so that the perpendicular  demagnetizing field cancels the magnetization contribution to $B$ ($H_D=-M$). This requires materials with large perpendicular anisotropy. 
A new dephasing mechanism associated with spin waves, specific to ferromagnets, has been studied. We have shown that it is easily tunable by a magnetic field.

%%%%%%%%%%%%%%%%%%%%%%%%%%%%%%%%%%%%%%%%%%%%%%%%%

G. T. is grateful to Ministry of Education, Culture, Sports, Science and
Technology, Japan and The Mitsubishi Foundation for financial support.
He thanks 
Laboratoire de Magn\'eisme Louis N\'{e}el for its hospitality
during his stay.

%%%%%%%%%%%%%%%%%%%%%%%%%%%%%%%%%%%%%%%%%%%%%%%%%%%%%%%%%%%%%%%%%%%%%

\end{document}